\documentclass[twocolumn,showpacs,preprintnumbers,amsmath,amssymb]{revtex4-1}

\usepackage{graphicx}
\usepackage{dcolumn}
\usepackage{bm}

\begin{document}

\preprint{condmat preprint}

\title{Using a tunable quantum wire to measure the large out-of-plane spin splitting of quasi two-dimensional holes in a GaAs nanostructure}

\author{A. Srinivasan, L. A. Yeoh, O. Klochan}
\author{T. P. Martin}
\email{Now at: Acoustics Division, Naval Research Laboratory, Washington, DC 20375, United States of America}
\author{J. C. H. Chen, A. P. Micolich}
\author{A. R. Hamilton}
\email{Alex.Hamilton@unsw.edu.au}
\affiliation{School of Physics, University of New South Wales, Sydney NSW 2052, Australia}
\author{D. Reuter and A. D. Wieck}
\affiliation{Angewandte Festkorperphysik, Ruhr-Universitat Bochum, D-44780 Bochum, Germany}

\date{\today}

\begin{abstract}
The out-of-plane $g$-factor $g^{*}_{\perp}$ for quasi-2D holes in a (100) GaAs heterostructure is studied using a variable width quantum wire.  A direct measurement of the Zeeman splitting is performed in a magnetic field applied perpendicular to the 2D plane. We measure an out-of-plane $g$-factor up to $g^{*}_{\perp} = 5$, which is larger than previous optical studies of $g^{*}_{\perp}$, and is approaching the long predicted but never experimentally verified out-of-plane $g$-factor of 7.2 for heavy holes.
\end{abstract}

\maketitle

The manipulation and control of the spin degree of freedom for representing and carrying information is one of the key ideas being put forward for electronic devices in the near future and the realisation of quantum computation~\cite{WolfSci01}.  Quantum confined holes in semiconductor nanostructures are interesting candidates for spintronics applications~\cite{DattaApl90}, due to the strong spin-orbit coupling in the valence band of direct-gap materials such as GaAs.  The p-like valence band states mean that the total angular momentum of holes \textit{J = l + s} = 3/2~\cite{WinklerBook03}.  One manifestation of the intriguing spin-3/2 nature of holes is a strong anisotropy in the transport properties of hole systems fabricated on GaAs/AlGaAs  heterostructures in the presence of an external magnetic field~\cite{WinklerPRL00, PapadakisPRL00, DanneauPRL06, KlochanNJP09}.

Confinement affects the orbital freedom of the carriers, which in the presence of strong spin-orbit coupling, will directly influence spin. For example, for two dimensionally (2D) confined heavy holes on the (100) GaAs surface, the Land\'{e} $g$-factor $g_{||}(k=0) = 0$ for in-plane directions, but $g_{\perp}(k=0) = 6\kappa = 7.2$ for the out-of-plane perpendicular direction~\cite{WinklerPRL00}, where $\kappa$ is a parameter originally introduced by Luttinger~\cite{LuttingerPR56}.  This anisotropy has been established theoretically for over two decades, and is known to occur as a result of the two dimensional confinement forcing the quantization axis for total angular momentum $J$ to point out of the 2D plane~\cite{WinklerBook03}.

Studies of the Zeeman splitting of holes in GaAs for in-plane directions have found a very small $g$-factor consistent with theory~\cite{DanneauPRL06, CsontosPRB07, DanneauPRL08, KoduvayurPRL08, KlochanNJP09, ChenNJP10}. However, despite a number of attempts over the past 20 years~\cite{KesterenPRB90, SapegaPRB92, GlasbergPRB01, SyperekPRL07, KuglerPRB09}, none have been able to experimentally replicate the large predicted out-of-plane $g$-factor of 7.2.  Moreover, there still exists a lack of consensus in the experimental data, with the measured $g^{*}_{\perp}$ varying from -0.7 to 2.9, meaning the theory has so far remained unsubstantiated. This discrepancy between theory and experiment is possibly due to the fact that optical studies are based upon measurements of bound excitons, where the lateral confinement prevents measurements in the 2D limit~\cite{SyperekPRL07, KuglerPRB09}.

An alternative method of measuring the $g$-factor in two dimensions is through transport measurements, using the tilted field Landau level coincidence method pioneered by Fang and Stiles~\cite{FangPR68}.  Whilst applicable to electron systems, a significant drawback to this approach is that it assumes an isotropic $g$-factor and a parabolic band structure, neither of which is the case for 2D holes.   Furthermore, this technique cannot show whether the Zeeman splitting is linear in $B$, as it is not a spectroscopic technique; it only measures the ratio of the Zeeman spitting to the cyclotron energy at a single value of $B$~\cite{YuanAPL09}.

In this paper, we circumvent the limitations of the Fang and Stiles method by directly measuring Zeeman splitting in a hole quantum wire to extract $g^{*}_{\perp}$. Confining the 2D holes to a one-dimensional (1D) channel makes it possible to perform energy spectroscopy and directly measure the $g$-factor at the 1D subband edges~\cite{DanneauPRL06, KlochanNJP09, ChenNJP10}.  The width of the channel can be tuned from the 1D limit where a single mode is occupied, up to the quasi 2D limit where a large number of modes are occupied.  The $g$-factor can thus be measured quantitatively between the 1D and 2D limits, for different magnetic field orientations.

The use of a magnetic field perpendicular to the plane of the quantum wire introduces complications due to the coupling to the orbital motion and the formation of Landau levels.  This limits measurements to low magnetic fields, making the spin-splitting small and hard to resolve.  We thus employ a novel method to measure Zeeman splitting for lower subbands:  The 1D subbands are first spin resolved in an in-plane field, then a small increasing out-of-plane field is applied whilst the in-plane component of the field is held constant. This allows the splitting to be easily resolved and quantified in small perpendicular fields below the quantum Hall regime.

\begin{figure}
\includegraphics{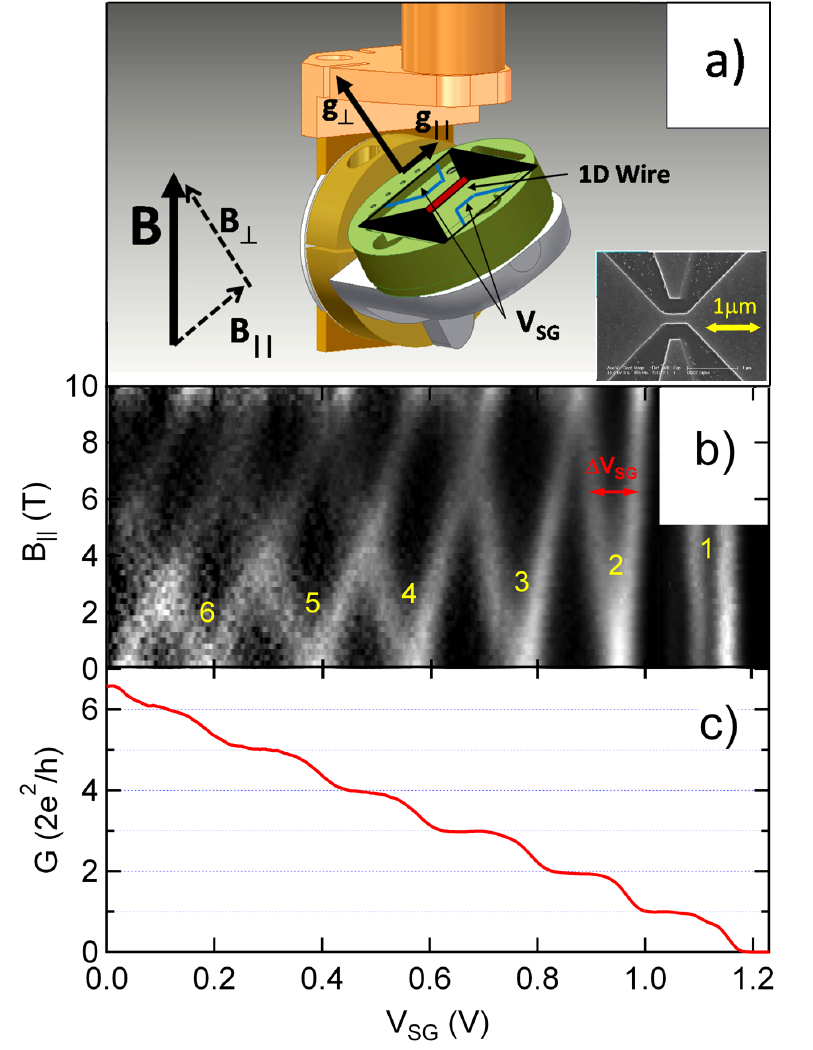}
\caption{(a) Schematic diagram of the sample mounted on an in-situ rotation system, showing the orientation of the field with respect to the wire. Inset: SEM image of quantum wire. (b) Zeeman splitting in an in-plane parallel magnetic field. A greyscale plot of the transconductance $\partial G/\partial V_{SG}$ is shown with dark regions corresponding to conductance plateaus (low transconductance) and bright regions corresponding to the risers between plateaus (high transconductance).  The superimposed numbers indicate the subband index. (c) The conductance G versus $V_{SG}$ measured at $B = 0$~T, which corresponds to a horizontal slice through the bottom of the greyscale plot in (b).}
\end{figure}


We now turn to our experiment.  The 1D wire was fabricated from a (100)-oriented heterostructure, in which 2D holes are induced at an Al$_{0.3}$Ga$_{0.7}$As/GaAs interface by applying a negative voltage (-0.7V) to a heavily doped cap layer. The peak 2D hole mobility was $\mu = 4.8 \times 10^5$~cm$^{2}$ V$^{-1}$s$^{-1}$ at a density $p=1.3 \times 10^{11}$~cm$^{-2}$ and temperature T = 100 mK. The 400 nm long and 200nm wide 1D channel, oriented along the [$01\overline{1}$] direction was defined by electron-beam lithography and shallow wet etching of the cap layer (see inset Fig.1(a)). Measurements were carried out in a dilution refrigerator using standard ac lock-in techniques at 17 Hz, with the sample mounted on an in-situ rotation system~\cite{YeohRSI10} (see Fig. 1(a)).

The Zeeman splitting was measured with the field applied in different orientations with respect to the quantum wire (see Fig. 1(a)): In-plane and parallel to the wire, out-of-plane and perpendicular to the wire, and finally, tilted to angles in between these two configurations. The effective g-factors for all measurements were calculated using a technique that combines Zeeman splitting in a magnetic field with splitting due to a source drain bias at $B=0$~\cite{PatelPRB91}. The Zeeman splitting of the 1D subbands is measured as a splitting in gate voltage $\Delta V_{SG}(B)$ from conductance traces.  A similar splitting in gate voltage, $\Delta V_{SG}(V_{SD})$, can also be induced with an applied d.c. source-drain bias, which separates the potentials of the source and drain by the bias energy $eV_{SD}$~\cite{MartinPRB10}. The lever arm, $\alpha = \partial V_{SD}/\partial V_{SG}$, extracted from source drain bias measurements allows one to convert the measured Zeeman splitting in gate voltage to an energy $\Delta E_{Z}(B)$. Full details of this measurement technique are given in the supplementary information~\cite{suppmat}.  We tuned the 1D channel width by applying a voltage $V_{SG}$ to two side gates, as shown in Fig. 1(a), and measured the $g$-factor versus 1D subband index $n$.

\begin{figure}
\includegraphics{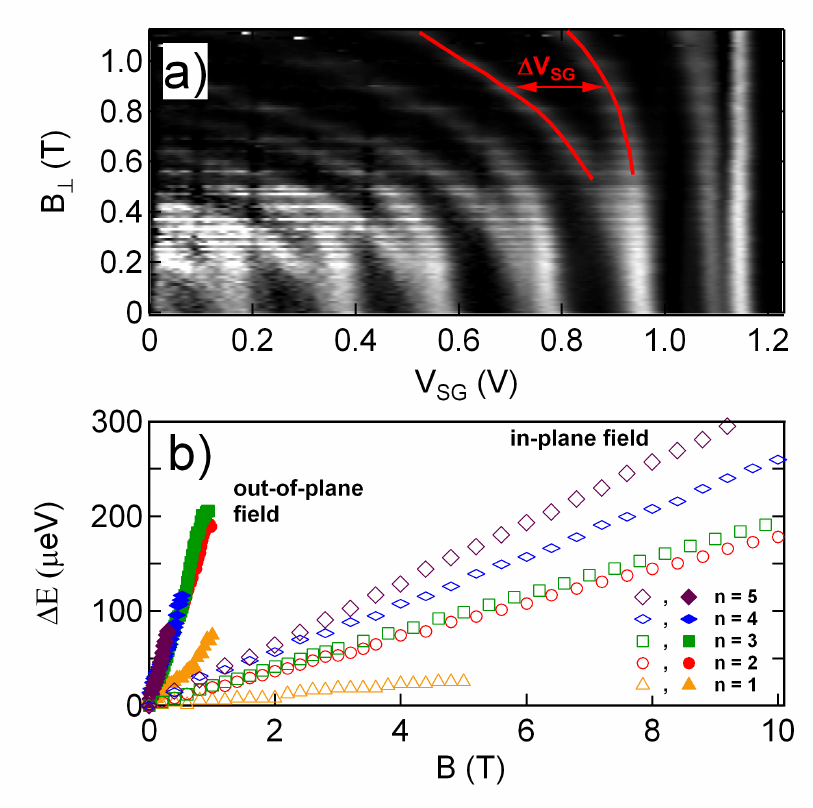}
\caption{(a) Zeeman splitting in an out-of-plane perpendicular magnetic field. A greyscale plot of the transconductance $\partial G/\partial V_{SG}$ is shown with magnetic field on the y-axis and $V_{SG}$ on the x-axis.  (b) The Zeeman energy splitting of each subband is plotted as a function of magnetic field for both field orientations.  The splitting is linear for both directions. Note: the Zeeman splitting of the first subband has been offset by $100~\mu$eV for clarity.}
\end{figure}


Measurements of the Zeeman splitting with the magnetic field applied in-plane and parallel to the quantum wire are shown in Fig. 1.  Figure 1(c) shows the clean conductance plateaus measured at $B_{||} = 0$~\cite{VanWeesPRL88,WharamJPC88}.  As the side gate voltage is made more positive, the hole channel is made narrower, and is eventually pinched off. At the widest point where six subbands are occupied, the electrical width of the quantum wire can be estimated using the known carrier density to calculate the Fermi wavelength.  We find width $\geq$ 180nm in good agreement with the lithographic width of 200nm. A greyscale plot of the transconductance $\partial G/\partial V_{SG}$, is presented versus magnetic field on the y-axis and $V_{SG}$ on the x-axis in Fig. 1(b).  The bright regions correspond to high transconductance, which are the risers between plateaus in the conductance in Fig. 1(c), hence marking the 1D subband edges. With the field aligned along the quantum wire, there is a strong linear Zeeman splitting of the 1D states with clear crossings between adjacent 1D subbands at higher fields.  The Zeeman splitting in gate voltage was measured directly from Fig. 1(b), and converted to an energy splitting $\Delta E_{Z}(B)$ (open data points in Fig. 2(b)) using the lever arm $\alpha$ ~\cite{suppmat}. 

We now extract the effective Land\'{e} g-factors $g_{||}^{*}$ for the in-plane direction using two methods. The Zeeman energy splitting of the n$^{\textrm{th}}$ degenerate subband is given by $\Delta E_{Z} = g^{*}_n \mu_B B$~\cite{DanneauPRL06}. Hence, the $g$-factor is simply the gradient of the Zeeman energy splitting in Fig. 2(b) multiplied by the factor $e/\mu_B$. The solid blue squares (Fig. 4) show $g_{||}^{*}$ obtained using this gradient method. In the second method, the field at which the spin down level of the n$^{\textrm{th}}$ subband crosses the spin up level of the (n+1)$^{\textrm{th}}$ subband (from Fig. 1(b)) is combined with the corresponding subband crossing in the d.c. source drain bias data. This gives an average effective $g$-factor for the n$^{\textrm{th}}$ and (n+1)$^{\textrm{th}}$ subbands, $<g_{n}^{*},g_{(n+1)}^{*}>$~\cite{PatelPRB91}. Values of $g_{||}^{*}$ calculated using this crossing method are shown by the blue open squares in Fig.4. The in-plane g-factors obtained from both methods are consistent with each other, and increase from 0.1 to 0.6 for n = 1 to 5. The $g$-factor saturates as the wire approaches the 2D limit (large n) consistent with previous studies for holes~\cite{ChenNJP10} and electrons~\cite{ThomasPRL96}.

We note that the first subband shows a kink at $0.7 \times 2e^{2}/h$, and an apparent spin-splitting even at B = 0. This 0.7 feature has been widely investigated ~\cite{ThomasPRL96,CronenPRL02}, although its origins are still debated~\cite{MicoJPC11}.  It is unlikely to be due to spin-orbit coupling effects since it is present even in systems with weak spin-orbit coupling such as n-type GaAs. Despite the fact that the first subband is already split at B = 0, the $g$-factor can still be extracted from the slope of $\Delta E_{Z}(B)$, consistent with the analysis for other subbands.

To extract the out-of-plane $g$-factor $g_{\perp}^{*}$, measurements of the Zeeman splitting were repeated with a magnetic field of up to 1T applied out-of-plane and perpendicular to the quantum wire, shown by the greyscale plot in Fig. 2(a).  This produces two key differences to the in-plane case.  Firstly, the Zeeman splitting is much larger, requiring much smaller fields to split the 1D subbands: In Fig. 2(a), the subbands are spin resolved well below 0.5 T, compared to an in-plane field $B_{||} > 2$~T needed for spin splitting. Secondly, there is a strong upward curvature in the 1D subbands for $B_{\perp}$. This is due to the additional confinement induced by the perpendicular field, causing the subbands to evolve into 2D Landau levels~\cite{BerggrenPRL86,WharamJPC88,VanWeesPRB88}.  At zero field, the subbands are purely determined by the electric confinement from $V_{SG}$. Once a perpendicular field is applied, hybrid magneto-electric subbands begin to form, and as the field is increased, the subbands move to higher energies, and therefore lower gate bias. Consequently, the analysis of Zeeman splitting is limited to low fields $B < 0.3~T$ to avoid entering the quantum Hall regime where the cyclotron diameter becomes small and comparable to the 1D confinement.

\begin{figure}
\includegraphics{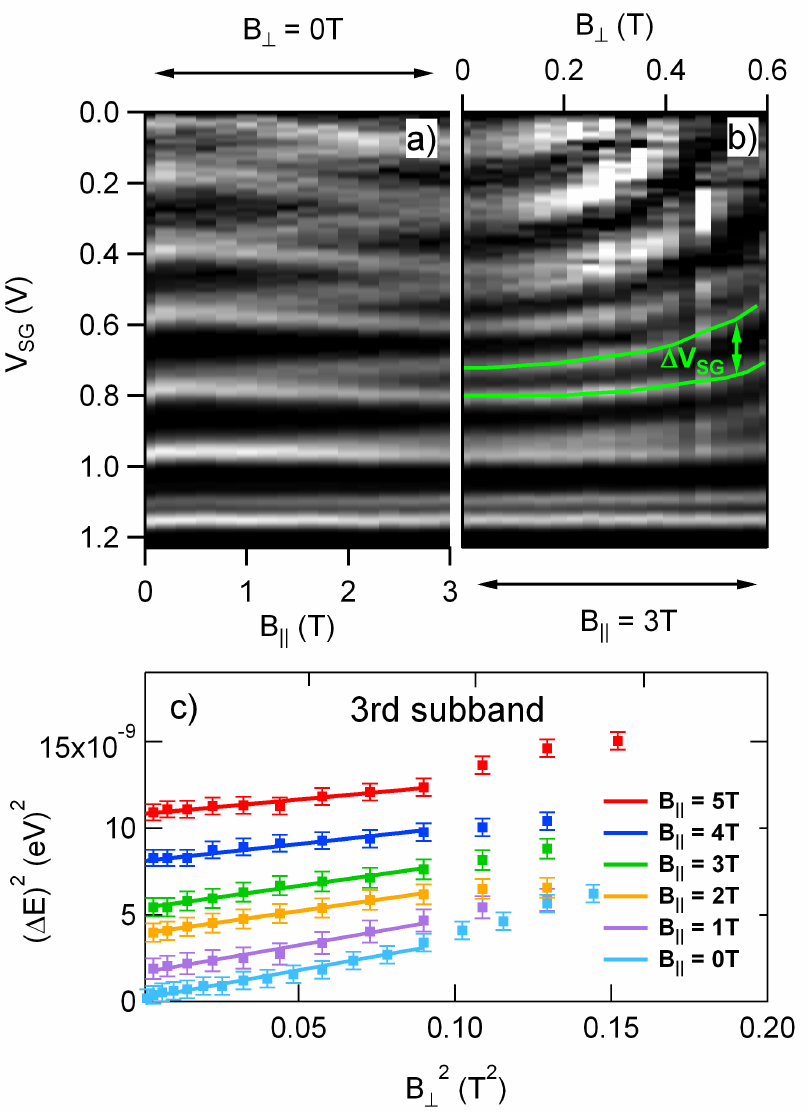}
\caption{(a) and (b): Greyscale plots of the Zeeman splitting in a tilted field: In (a), a 3 T in-plane magnetic field is first applied parallel to the wire direction to spin resolve the subbands. Then in (b), the sample is tilted ($0^{o}$ to $10^{o}$) into the perpendicular plane whilst the parallel component of the field is kept constant. (c) Perpendicular field Zeeman splitting rates calculated by fitting points to subband 3. This was carried out for 6 different tilted field measurements ($B_{||} = 0 - 5$~T).  The green points ($B_{||} = 3$~T) correspond to data from Fig. 3(b); the light blue points ($B_{||} = 0$) correspond to data from Fig. 2(a).}
\end{figure}

The splitting in gate voltage was again converted to an energy splitting $\Delta E_{Z}(B)$, shown by the solid data points in Fig. 2(b).  For an in-plane field, the subbands split but remain centred at the same $V_{SG}$. However, in an out-of-plane field, the levels split but also become shifted to lower $V_{SG}$.  Since the lever-arm is gate voltage dependent, this must be accounted for when converting $\Delta V_{SG}(B)$ to $\Delta E_{Z}(B)$~\cite{suppmat}.  It is found that the measured Zeeman splitting in an out-of-plane field is also linear despite the curvature of the subbands. This allows an estimate of the effective out-of-plane $g$-factor from the gradient of the data in Fig. 2(b) for $B_{\perp} < 0.3$~T.  This was carried out for the 1st, 3rd, 4th and 5th subbands from the data in Fig. 2 (the Zeeman splitting of subband 2 could not be clearly resolved below 0.3 T). To check that the results are not sensitive to cut-off, $g_{\perp}^{*}$ was also calculated using only data up to $B_{\perp} = 0.2$~T, and we find that the obtained g-factors agree within experimental error with those calculated for $B_{\perp} < 0.3$~T. The extracted values of the out-of-plane $g$-factor are shown by the open red squares in Fig. 4. It is clear from the splitting in Fig. 2(b) and the points in Fig. 4 that $g_{\perp}^{*}$ is an order of magnitude larger than $g_{||}^{*}$, consistent with theoretical predictions.

We use a novel technique to extract $g_{\perp}^{*}$ for the lower subbands: We first completely resolve the spin with an in-plane field aligned along the quantum wire (e.g. $B_{||} = 3$~T as in Fig. 3(a)).  The out-of-plane component of the field $B_{\perp}$ is then increased in increments of 0.05 T, while $B_{||}$ is kept constant so any additional splitting is due to $B_{\perp}$ only (Fig. 3(b)). This is achieved using a rotation mechanism enabling the sample to remain below 200mK during rotation~\cite{YeohRSI10}.

Tilting the sample in this way allows us to resolve the out-of-plane spin splitting for subband 2 as well as subband 3 in the low field limit ($B_{\perp} < 0.3$~T). This measurement was repeated for 5 different values of parallel field between 1-5 T providing multiple datasets to obtain an average of the Zeeman splitting rates for both subbands.

The use of tilted fields makes the analysis and subsequent calculation of $g_{\perp}^{*}$ more complex, as the Zeeman splitting is dependent on the total field. The total $g$-factor can be expressed in terms of its individual in-plane and out-of-plane components~\cite{KuglerPRB09}.  When the sample is subject to a tilted field as in Fig.3 (b), the total Zeeman splitting $\Delta E_{TOT}$ is given by: 
$\Delta E_{TOT}^2~=~{(g_{\perp}\mu_{B}B_{\perp})}^2 + {(g_{||}\mu_{B}B_{||})}^2$.  Hence, plotting $(\Delta E_{Z}(B))^{2}$ versus $B_{\perp}^{2}$ will result in a linear relationship with a gradient proportional to $g_{\perp}^{2}$, as observed in Fig. 3 (c).  This data is analysed only for the 2$^{\textrm{nd}}$ and 3$^{\textrm{rd}}$ subbands, as the higher subbands rapidly merge together and cross, so that spin-splitting cannot be clearly resolved. The square of the energy splitting was measured for each $B_{||}$ dataset, as shown in Fig.3 (c) for subband 3.  We observe a linear Zeeman splitting as a function of out-of-plane field for $B_{\perp} < 0.3$~T, which allows the calculation of $g_{\perp}^{*}$ for the 2$^{\textrm{nd}}$ and 3$^{\textrm{rd}}$ subbands.

The $g_{\perp}^{*}$ values extracted from these tilted field measurements, averaged over five $B_{||}$ datasets (1-5T), are shown by the red solid circles in Fig. 4.  $g_{\perp}^{*}$ obtained for subband 3 shows good agreement between the fully perpendicular field and tilted field measurements, validating the tilted field measurement technique.

\begin{figure}
\includegraphics{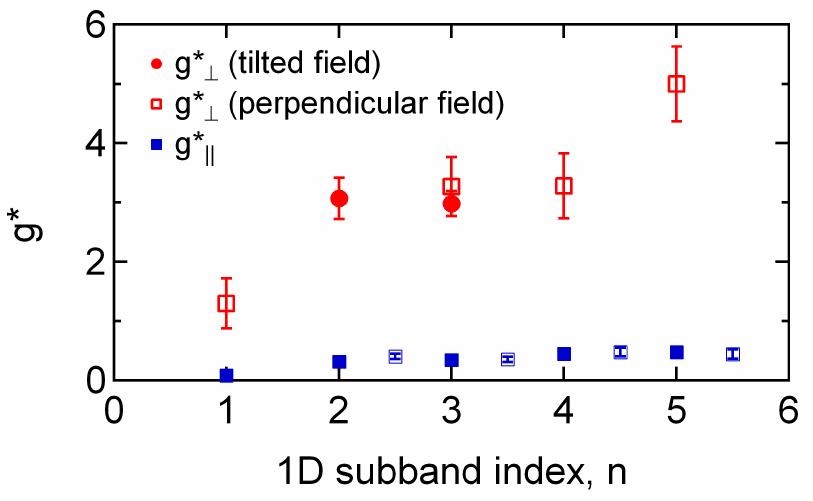}
\caption{The effective Land\'{e} $g$-factor $g_{n}^{*}$ is plotted as a function of 1D subband index n.  The in-plane $g$-factors $g_{||}^{*}$ were calculated using both the crossing method (open blue squares) and gradient method (solid blue squares) from the Zeeman splitting measurements in Fig.1 (a).  $g_{\perp}^{*}$ was calculated from both the out-of-plane perpendicular Zeeman splitting measurements in Fig.2 (a) (for subbands 1,3,4 and 5 shown by open red squares), and from the tilted field Zeeman splitting measurements in Fig.3 (b) (for subbands 2 and 3 shown by solid red circles).}
\end{figure}

We find that the effective Land\'{e} g-factors measured for the out-of-plane direction are significantly larger than the in-plane g-factors: The anisotropy of the spin splitting $g_{\perp}^{*} / g_{||}^{*}$ exceeds 8.5.  This large anisotropy agrees with the theoretical prediction that the natural quantization axis for 2D heavy holes is pointed out-of-plane~\cite{WinklerBook03}.
The out-of-plane $g$-factor $g_{\perp}^{*}$ appears to grow larger with increasing subband index, as the quasi 1D system becomes more two-dimensional. This may be explained by the natural quantization axis reorienting from out-of-plane in the 2D limit to along the wire in the 1D limit~\cite{ChenNJP10}.  Since the quantization axis is perpendicular to B in the 1D limit, the spin-splitting of heavy hole states is suppressed for low subband index.  Furthermore, for the lower subbands which correspond to stronger confinement, we observe $g_{\perp}^{*} < 3$, consistent with the optical studies of highly confined excitons~\cite{KesterenPRB90, SapegaPRB92, GlasbergPRB01, SyperekPRL07, KuglerPRB09}.  When the channel is much wider at n = 5, we find that $g_{\perp}^{*} \simeq 5$, approaching the theoretical 2D limit of 7.2~\cite{WinklerPRL00}. Higher subbands cannot be resolved with our device, but we note that in electron quantum point contacts, the system draws near the bulk $g$-factor by the 10$^{\textrm{th}}$ 1D subband~\cite{ThomasPRL96}.  Our measured trend is also qualitatively consistent with recent theoretical work in which the g-factor in a hole quantum wire was found to increase as the wire becomes more 2D-like~\cite{MagPRB11}.

In the case of the hole in-plane $g$-factor, there is again an increase in $g_{||}^{*}$ with subband index, however it is of a much smaller magnitude. This trend is in agreement with results published earlier~\cite{ChenNJP10}, and may be due to the dependence of $g_{||}^{*}$ on the wave vector $k_{n}$.

In 1D electron systems it is well established that the g-factor increases as the system becomes more one-dimensional, due to the increasing importance of exchange and correlation effects at low densities~\cite{ThomasPRL96}.  In contrast, for the hole system studied here, the g-factor decrease as the system is made more one dimensional, which suggests that there is no exchange enhancement for [100] heavy holes in GaAs.  It is interesting to note that there are also suggestions that there is an absence of exchange enhancement in 2D hole systems, based on a comparison of experimental and theoretical analyses of subband depopulation by in-plane fields ~\cite{WinklerPRL00}.

In summary, we have conducted a direct measurement of Zeeman splitting for heavy holes in GaAs by performing energy spectroscopy on a quasi 1D quantum wire. We have also developed a new method for studying out-of-plane Zeeman splitting by resolving the spin subbands in an in-plane field before independently adding an out-of-plane field.  A linear Zeeman splitting was observed in the low out-of-plane field limit $B_{\perp} < 0.3$~T, from which we extract $g_{\perp}^{*}$ of up to 5.  Our measurement goes some way towards resolving the long standing discrepancy between theory and experiment for the out-of-plane $g$-factor of heavy holes in GaAs.  These results provide a bridge between the optical experiments of strongly confined excitons~\cite{KesterenPRB90, SapegaPRB92, GlasbergPRB01, SyperekPRL07, KuglerPRB09} and the predicted theory for purely two-dimensional holes.


We wish to acknowledge J. Cochrane for technical support and thank U. Z\"{u}licke and T. Li for discussions.  This work was supported by the Australian Research Council (ARC) under the DP scheme. APM and ARH acknowledge ARC Future and Professorial Fellowships, respectively.  DR and ADW acknowledge support from DFG SPP1285 and BMBF QuaHL-Rep 01BQ1035.


Supporting Information Available: Full details of the Zeeman energy splitting measurement technique is given in the supplemetary information.


\begin{thebibliography}:

\bibitem{WolfSci01} S.A. Wolf, D.D. Awschalom, R.A. Buhrman, J.M. Daughton, S. von Molnar, M.L. Roukes, A.Y. Chtchelkanova, and D.M. Treger, \textit{Science} {\bf 294}, 1488 (2001).

\bibitem{DattaApl90} S. Datta and B. Das, \textit{Appl. Phys. Lett.} {\bf 56}, 665 (1990).

\bibitem{WinklerBook03} R. Winkler, {\textit{Spin-orbit coupling effects in two-dimensional electron and hole systems}, (Springer Tracts in Modern Physics, Vol. 191, Springer, Berlin, 2003).

\bibitem{WinklerPRL00} R. Winkler, S. J. Papadakis, E. P. De Poortere, and M. Shayegan, \textit{Phys. Rev. Lett.} {\bf 85}, 4574 (2000).

\bibitem{PapadakisPRL00} S.J. Papadakis, E.P. De Poortere, M. Shayegan and R. Winkler, \textit{Phys. Rev. Lett.} {\bf 84}, 5592-5595 (2000).

\bibitem{DanneauPRL06} R. Danneau, O. Klochan, W.R. Clarke, L.H. Ho, A.P. Micolich, M.Y. Simmons, A.R. Hamilton, M. Pepper, D.A. Ritchie and U. Z\"{u}licke, \textit{Phys. Rev. Lett.} {\bf 97}, 026403 (2006).

\bibitem{KlochanNJP09} O. Klochan, A.P. Micolich, L.H. Ho, A.R. Hamilton, K. Muraki and Y. Hirayama, \textit{New. J. Phys.} {\bf 11}, 043018 (2009).

\bibitem{LuttingerPR56} J.M. Luttinger \textit{Phys. Rev.} \textbf{102}, 1030–1041 (1956).

\bibitem{ChenNJP10} J.C.H Chen, O. Klochan, A.P. Micolich, A.R. Hamilton, T.P. Martin, L.H. Ho, U. Z\"{u}licke, D. Reuter and A.D. Wieck, \textit{New. J. Phys.} {\bf 12}, 033043 (2010).

\bibitem{CsontosPRB07} D. Csontos and U. Z\"{u}licke, \textit{Phys. Rev. B} {\bf 76}, 073313 (2007).

\bibitem{DanneauPRL08} R. Danneau, O. Klochan, W.R. Clarke, L.H. Ho, A.P. Micolich, M.Y. Simmons, A.R. Hamilton, M. Pepper, and D.A. Ritchie,  \textit{Phys. Rev. Lett.} {\bf 100}, 016403 (2008).

\bibitem{KoduvayurPRL08} S.P. Koduvayur, L.P. Rokhinson, D.C. Tsui, L.N. Pfeiffer and K.W. West,  \textit{Phys. Rev. Lett.} \textbf{100}, 126401 (2008).

\bibitem{GlasbergPRB01} S. Glasberg, H. Shtrikman and I. Bar-Joseph, \textit{Phys. Rev. B.} \textbf{63}, 210308(R) (2001).

\bibitem{KesterenPRB90} H.W. van Kesteren, E.C. Cosman, W.A.J.A. van der Poel and C.T. Foxon, \textit{Phys. Rev. B.} \textbf{41}, 5283-5292 (1990).

\bibitem{KuglerPRB09} M. Kugler, T. Andlauer, T. Korn, A. Wagner, S. Fehringer, R. Schulz, M. Kubov\'{a}, C. Gerl, D. Schuh, W. Wegscheider, P. Vogl, and C. Schuller, \textit{Phys. Rev. B.} \textbf{80}, 035325 (2009).

\bibitem{SyperekPRL07} M. Syperek, D.R. Yakovlev, A. Greilich, J. Misiewicz, M. Bayer, D. Reuter, and A.D. Wieck, \textit{Phys. Rev. Lett.} \textbf{99}, 187401 (2007).

\bibitem{SapegaPRB92} V.F. Sapega, M. Cardona, K. Ploog, E.L. Ivchenko and D.N. Mirlin,  \textit{Phys. Rev. B.} \textbf{45}, 4320–4326 (1992).

\bibitem{FangPR68} F.F. Fang and P.J. Stiles, \textit{Phys. Rev.} \textbf{174}, 823 (1968).

\bibitem{YuanAPL09} Z.Q. Yuan, R.R. Du, M.J. Manfra, L.N. Pfeiffer and K.W. West, \textit{Appl. Phys. Lett.} \textbf{94}, 052103 (2009).

\bibitem{YeohRSI10}  L.A. Yeoh, A. Srinivasan, T.P. Martin, O. Klochan, A.P. Micolich and A.R. Hamilton,  \textit{Rev. Sci. Instrum.} \textbf{81}, 113905 (2010).

\bibitem{PatelPRB91} N.K. Patel, J.T. Nicholls, L. Martin-Moreno, M. Pepper, J.E.F. Frost, D.A. Ritchie and G.A.C. Jones, \textit{Phys. Rev. B.} \textbf{44}, 13549 (1991).

\bibitem{MartinPRB10} T.P. Martin, A. Szorkovszky, A.P. Micolich, A.R. Hamilton, C.A. Marlow, R.P. Taylor, H. Linke and
H.Q. Xu, \textit{Phys. Rev. B.} \textbf{81}, 041303(R) (2010).

\bibitem{suppmat} See supplementary information for a detailed description of the Zeeman splitting measurement technique.

\bibitem{VanWeesPRL88} B.J. van Wees, H. van Houten, C.W.J. Beenakker, J.G. Williamson, L.P. Kouwenhoven, D. van der Marel, and C.T. Foxon, \textit{Phys. Rev. Lett.} \textbf{60}, 848 (1988).

\bibitem{WharamJPC88} D.A. Wharam, T.J. Thornton, R. Newbury, M. Pepper, H. Ahmed, J.E.F. Frost, D.G. Hasko, D.C. Peacock, D.A. Ritchie and G.A.C. Jones, \textit{J. Phys. C: Solid State Phys.} \textbf{21}, L209-L214 (1988).

\bibitem{ThomasPRL96} K.J. Thomas, J.T. Nicholls, M.Y. Simmons, M. Pepper, D.R. Mace and D.A. Ritchie,  \textit{Phys. Rev. Lett.} {\bf 77}, 135-138 (1996).

\bibitem{CronenPRL02} S.M. Cronenwett et al., \textit{Phys. Rev. Lett.} {\bf 88}, 226805 (2002).

\bibitem{MicoJPC11} A.P. Micolich, \textit{J. Phys.: Condens. Matter} \textbf{23}, 443201 (2011). 

\bibitem{BerggrenPRL86} K.F. Berggren, T.J. Thornton, D.J. Newson and M. Pepper \textit{Phys. Rev. Lett.} \textbf{57}, 1769 (1986).

\bibitem{VanWeesPRB88} B.J. van Wees, L.P. Kouwenhoven, H. van Houten, C.W.J. Beenakker, J.E. Mooij, C.T. Foxon and J. J. Harris, \textit{Phys. Rev. B.} \textbf{38}, 3625-3627 (1988).

\bibitem{MagPRB11} M. M. Gelabert and L. Serra \textit{Phys. Rev. B.} \textbf{84}, 075343 (2011).

}\end{thebibliography}
\end{document}